\documentstyle[prl,aps,12pt,preprint]{revtex}
\title{Hot Electron Magnetotransport in a Spin-Valve Transistor at Finite temperatures }
\author{Jisang Hong}
\address{Max-Planck-Institut f{\"u}r Mikrostrukturphysik, Weinberg 2, D-06120 Halle, Germany}
\begin{document}
\maketitle
\begin{abstract}
The hot electron  magnetotransport in a spin-valve transistor has been theoretically explored at finite  temperatures. We have explored the parallel and anti-parallel collector current changing the relative spin orientation of the ferromagnetic layers at finite temperatures. In this model calculations, hot electron energy redistribution due to spatial inhomogeneity of Schottky barrier heights and hot electron spin polarization in the ferromagnetic layer at finite temperatures have been taken into account. The results of this model calculations accord with the experimental data semi-quantitative manner. We therefore suggest that both effects remarked above should be taken into account substantially when one explores the hot electron magnetotransport in a spin-valve system transistor at finite temperatures.
\end{abstract}
\newpage
\newcommand{\eb}{\begin{eqnarray}}
\newcommand{\ee}{\end{eqnarray}}
\section*{}
Since the discovery of the giant magneto resistance (GMR) \cite{GMR} in mganetic multilayer structure, the spin dependent transport has been extensively studied because of the fundamental interests as well as practical purposes. For instance, magnetic tunneling junction (MTJ) \cite{junction} has been explored very actively for real device applications. Interestingly, Monsma {\it et al} presented a spin-valve transistor \cite{valve} as a new type of magnetoelectronic device. One major difference between the MTJ and spin-valve transistor is the transport property of electrons. One needs to explore the {\it hot} electron transport in the spin-valve transistor, while the Fermi electrons contribute to the current in the MTJ. The hot electron transport property is related to the unoccupied density of states above the Fermi level, and has an exponential dependence on the inelastic mean free path \cite{exponential}. Besides, the spin-valve transistor has a different structure \cite{structure} from the MTJ. Very recently, Jansen {\it et al} \cite{Jansen} reported the temperature dependence of the collector current changing the relative spin orientation of the magnetic moments in the ferromagnetic layers as well as the magnetocurrent. According to their observation, when the magnetic moments are parallel the collector current (parallel collector current) is increasing up to 200 K and decreasing above that, while the anti-parallel collector current is increasing up to room temperature. In addition, a huge magnetocurrent is also observed (roughly 350 \% at room temperature). Spin mixing mechanism due to thermal spin waves and energy redistribution due to spatial inhomogeneity of Schottky barrier heights at finite temperatures are suggested to account for the experimental data.

Regarding the issue of spin mixing effect, a theoretical calculations \cite{theoretical} has been presented to explore the relative importance of spin mixing  and hot electron spin polarization  at finite temperatures. Interestingly, the theoretical calculations suggest that the hot electron spin polarization has a substantial contribution to the hot electron magnetotransport, and this suggestion is supported by the magnetocurrent at finite temperatures. However, the theoretical calculations \cite{theoretical} can not account for the parallel and anti-parallel collector current at finite temperatures because only the relative importance of hot electron spin polarization and spin mixing due to thermal spin waves has been explored. In this model calculations, we shall study the hot electron magnetotransport taking into account the spatial inhomogeneity of Schottky barrier height distribution and spin dependent self energy effect in the ferromagnetic layers. 

The spin-valve transistor has typically $Si/N/F/N/F/N/Si$ structure \cite{structure} where $N$ stands for the normal metal layer and $F$ for the ferromagnetic layer . Then, the electrons injected across the Schottky barrier at the emitter side penetrate the spin-valve base, and the energy of injected hot electrons is influenced by the distribution of Schottky barrier heights \cite{barrier}. Once the electrons start to penetrate the spin-valve base we need to explore the Green's function $G_\sigma(\vec{k},E)$, which describes the propagation of the electrons of spin $\sigma$ in each layer. We can write this as
\eb
G_\sigma(\vec{k},E)=\frac{1}{E-\epsilon_\sigma(\vec{k})-\Sigma_\sigma(\vec{k},E)}.
\ee
In the normal metal layer the Green's function has no spin dependence, so that the hot electrons are not spin polarized until they reach the first ferromagnetic layer. However, in the ferromagnetic material the hot electron has strong spin dependent self energy \cite{self}. Thus, the inelastic mean free path of the hot electron is spin dependent, which results in the spin dependent attenuation in the ferromagnet. Then, by the virtue of the fact that the self energy has spin dependence, the hot electrons will be spin polarized after passing the ferromagnetic layer. Of course, the hot electrons will be attenuated in the normal metal layer as well. Since the hot electron transport has an exponential dependence on the inelastic mean free path \cite{exponential}, and  the attenuation in the normal metal layer has no influence on the spin dependent hot electron magnetotransport, we are able to focus our interests on the hot electron transport in the ferromagnetic layers.

In this model calculations it is assumed that we have the same type of normal metal layers with the same thickness in the spin-valve base. We then define $\Gamma_N=exp(-w_N/l_N)$ to account for the attenuation in the normal metal layer where $w_N$ is the thickness of the normal metal layer, and $l_N$ is the inelastic mean free path in the normal metal layer. As remarked above, the hot electron has strong spin dependent inelastic mean free path in the ferromagnets we therefore define $\gamma_{M(m)_i}=exp(-w_i/l_{M(m)_i})$ to consider the attenuation in the ferromagnetic layer $F_i$ of majority (minority) spin electrons, respectively. Here, the $w_i$ is the thickness of the ferromagnetic layer, and the $l_{M(m)_i}$ stands for the inelastic mean free path of majority (minority) spin electron. Generally speaking, the hot electron inelastic mean free path depends on the energy and temperature, hence we need to take into account those dependence for quantitative analysis of the hot electron magnetotransport. 

Taking into account the distribution of Schottky barrier heights and spin dependent self energy effect, we can write the parallel collector current 
\eb
\tilde{I}^P(T)&=&\int^{\epsilon_u}_{\epsilon_l}d\epsilon \int^T_0 dT^\prime D(\tilde{\epsilon}(T^\prime))\Gamma^3_N(\tilde{\epsilon}(T^\prime))\gamma_{M_1}(\tilde{\epsilon}(T^\prime))\gamma_{M_2}(\tilde{\epsilon}(T^\prime)) \nonumber \\
&&\times [1+\frac{\gamma_{m_1}(\tilde{\epsilon(T^\prime))}}{\gamma_{M_1}(\tilde{\epsilon(T^\prime))}}\frac{\gamma_{m_2}(\tilde{\epsilon(T^\prime))}}{\gamma_{M_2}(\tilde{\epsilon(T^\prime))}} ]\Theta(\tilde{\epsilon}(T^\prime)-V_b),
\ee
and the anti-parallel collector current is
\eb
\tilde{I}^{AP}(T)&=&\int^{\epsilon_u}_{\epsilon_l}d\epsilon \int^T_0 dT^\prime D(\tilde{\epsilon}(T^\prime))\Gamma^3_N(\tilde{\epsilon}(T^\prime))\gamma_{M_1}(\tilde{\epsilon}(T^\prime))\gamma_{M_2}(\tilde{\epsilon}(T^\prime)) \nonumber \\
&&\times [\frac{\gamma_{m_1}(\tilde{\epsilon(T^\prime))}}{\gamma_{M_1}(\tilde{\epsilon(T^\prime))}}+\frac{\gamma_{m_2}(\tilde{\epsilon(T^\prime))}}{\gamma_{M_2}(\tilde{\epsilon(T^\prime))}} ]\Theta(\tilde{\epsilon}(T^\prime)-V_b)
\ee
where $\Theta$ is a step function. The $V_b$ is the Schottky barrier height at the collector side, and the $\tilde{\epsilon}(T^\prime)$ is the energy of hot electron. In what follows, the energy is measured from the Fermi level of the metallic base. As mentioned above the Schottky barrier heights are not constant, but have spatial distribution \cite {barrier}, and this affects the energy distribution of injected hot electrons. We then denote the $\epsilon_u$ and $\epsilon_l$ as the upper and lower bound of the hot electron energy at zero temperature. At finite temperature T  this electron gains a fraction energy due to thermal effect with the $4k_BT$ width \cite{Jansen,width}, then the energy of the electron at finite temperatures can be written as 
\eb
\tilde{\epsilon}(T^\prime)=\epsilon+4k_BT^\prime
\ee
where $\epsilon$ is the energy at zero temperature. The function $D$ displays the energy distribution at finite temperatures. Based on the Schottky barrier heights measurement \cite{barrier}, in this calculations we assume that the energy of hot electrons has Gaussian distribution at zero temperature. At finite temperatures, the energy of hot electrons will be redistributed due to thermal energy. We thus model the energy distribution of the hot electrons at finite temperatures as   
\eb
D(\tilde{\epsilon}(T^\prime))=\frac{N_0}{2}c_1exp[-\alpha_1(\epsilon-\epsilon_m)^2] \times c_2exp[-\alpha_2(T^\prime/T)]
\ee
where the $c_1$ and $c_2$ are the normalization constants, $\epsilon_m$ is the energy of the maximum distribution at zero temperature, $\alpha_1$ and $\alpha_2$ describe the width of the distribution, and $N_0$ is the total number of injected electron (spin up and spin down) across the Schottky barrier per unit time per unit area. In the Eqs. (2) and (3), we can replace the ratio of the spin dependent attenuation in the ferromagnetic layer by hot electron spin polarization $P_{H_i}(\tilde{\epsilon}(T^\prime)$ using the relation 
\eb
\frac{\gamma_{m_i}(\tilde{\epsilon}(T^\prime))}{\gamma_{M_i}(\tilde{\epsilon}(T^\prime))}=\frac{1-P_{H_i}(\tilde{\epsilon}(T^\prime)}{1+P_{H_i}(\tilde{\epsilon}(T^\prime)} 
\ee
We thus obtain the expression for the parallel and anti-parallel collector current  
\eb
\tilde{I}^P(T)&=&\int^{\epsilon_u}_{\epsilon_l}d\epsilon \int^T_0 dT^\prime D(\tilde{\epsilon}(T^\prime))\Gamma^3_N(\tilde{\epsilon}(T^\prime))g_1(\tilde{\epsilon}(T^\prime))g_2(\tilde{\epsilon}(T^\prime))\Theta(\tilde{\epsilon}(T^\prime)-V_b) \nonumber \\
&&\times (1+P_{H_1}(\tilde{\epsilon}(T^\prime)))(1+P_{H_2}(\tilde{\epsilon}(T^\prime)))[1+\frac{1-P_{H_1}(\tilde{\epsilon}(T^\prime))}{1+P_{H_1}(\tilde{\epsilon}(T^\prime))}\frac{1-P_{H_2}(\tilde{\epsilon}(T^\prime))}{1+P_{H_2}(\tilde{\epsilon}(T^\prime))},
\ee 
\eb
\tilde{I}^{AP}(T)&=&\int^{\epsilon_u}_{\epsilon_l}d\epsilon \int^T_0 dT^\prime D(\tilde{\epsilon}(T^\prime))\Gamma^3_N(\tilde{\epsilon}(T^\prime))g_1(\tilde{\epsilon}(T^\prime))g_2(\tilde{\epsilon}(T^\prime))\Theta(\tilde{\epsilon}(T^\prime)-V_b) \nonumber \\
&&\times (1+P_{H_1}(\tilde{\epsilon}(T^\prime)))(1+P_{H_2}(\tilde{\epsilon}(T^\prime)))[\frac{1-P_{H_1}(\tilde{\epsilon}(T^\prime))}{1+P_{H_1}(\tilde{\epsilon}(T^\prime))}+\frac{1-P_{H_2}(\tilde{\epsilon}(T^\prime))}{1+P_{H_2}(\tilde{\epsilon}(T^\prime))},
\ee
where $g_i(\tilde{\epsilon}(T^\prime))$ is a spin averaged attenuation in the ferromagnetic layer, so that it has no spin dependence. One can understand this from the Eq. (6). To analyze the hot electron magnetotransport, it is necessary to know the temperature and energy dependence of the hot electron spin polarization and inelastic mean free path. In the discussion of these issues, the hot electron inelastic mean free path varying the spin and energy in the ferromagnets \cite{self} have been presented at zero temperature. In this theoretical calculations, various spin dependent inelastic scattering processes have been included such as spin wave excitations, Stoner excitations, and spin non-flip processes. However, the temperature dependence of hot electron inelastic mean free path and spin polarization at low energy has not been explored extensively neither experimentally nor theoretically to my best knowledge so far, although there is an example of lifetime measurement of Co \cite{Co} (the experimental data do not contain the information of the temperature dependence). We will therefore take advantage of zero temperature calculations \cite{self} and model the hot electron spin polarization at finite temperatures. For the $\Gamma_N({\tilde \epsilon(T^\prime}))$, it is of importance to note that the attenuation of low energy electron in the normal metal is around 100 $\AA$ \cite{normal}, It is several times greater than that calculated in the ferromagnets \cite{self}. This may imply that the inelastic scattering in the ferromagnetic layers enters importantly into the hot electron magnetotransport. Hence, we assume that the inelastic mean free path in normal metal layer is constant within the temperature and energy ranges of our interest. We also replace the $\tilde g_i({\epsilon(T^\prime)})$ by $\tilde g_i({\epsilon(0)})$, and one can note that those terms enter into the parallel and anti-parallel collector simultaneously. Hence they have no influence on the spin dependent magnetotransport, save for the magnitude of the collector current. This property enables us to explore the parallel and anti-parallel collector current expressed below since our interests are the magnetotransport depending on the relative spin configuration and magnetocurrent. Thus, we will explore the collector current which will be the maximum magnitude of the collector current in the spin-valve transistor
\eb
I^P(T)&=&\int^{\epsilon_u}_{\epsilon_l}d\epsilon \int^T_0 dT^\prime D(\tilde{\epsilon}(T^\prime))\Gamma^3_N(\tilde{\epsilon}(0))g_1(\tilde{\epsilon}(0))g_2(\tilde{\epsilon}(0))\Theta(\tilde{\epsilon}(T^\prime)-V_b) \nonumber \\
&&\times (1+P_{H_1}(\tilde{\epsilon}(T^\prime)))(1+P_{H_2}(\tilde{\epsilon}(T^\prime)))[1+\frac{1-P_{H_1}(\tilde{\epsilon}(T^\prime))}{1+P_{H_1}(\tilde{\epsilon}(T^\prime))}\frac{1-P_{H_2}(\tilde{\epsilon}(T^\prime))}{1+P_{H_2}(\tilde{\epsilon}(T^\prime))},
\ee
and
\eb
I^{AP}(T)&=&\int^{\epsilon_u}_{\epsilon_l}d\epsilon \int^T_0 dT^\prime D(\tilde{\epsilon}(T^\prime))\Gamma^3_N(\tilde{\epsilon}(0))g_1(\tilde{\epsilon}(0))g_2(\tilde{\epsilon}(0))\Theta(\tilde{\epsilon}(T^\prime)-V_b) \nonumber \\
&&\times (1+P_{H_1}(\tilde{\epsilon}(T^\prime)))(1+P_{H_2}(\tilde{\epsilon}(T^\prime)))[\frac{1-P_{H_1}(\tilde{\epsilon}(T^\prime))}{1+P_{H_1}(\tilde{\epsilon}(T^\prime))}+\frac{1-P_{H_2}(\tilde{\epsilon}(T^\prime))}{1+P_{H_2}(\tilde{\epsilon}(T^\prime))}.
\ee
One can also easily obtain the magnetocurrent by the definition
\eb
MC(T)=\frac{I^P(T)-I^{AP}(T)}{I^{AP}(T)}
\ee

In this model calculations we take the inelastic mean free path in the normal metal as $90 \AA$, and for the functions $g_i$ and $P_H$ in Eqs. (9) and (10) we will adopt the results presented in Ref. \cite{self}. To mimic the structure reported in the Ref. \cite{Jansen}, the thickness of a normal metal layer has been taken as 35 $\AA$. The first ferromagnetic layer is assumed be as $Ni$ with thickness 60 $\AA$ and $Fe$ for the second ferromagnetic layer with thickness 30 $\AA$. Since the Schottky barrier heights have roughly 0.2 eV distribution \cite{barrier}, we take 0.8 eV, 0.9 eV, 1 eV  for $\epsilon_l$, $\epsilon_m$, $\epsilon_u$ in Eqs. (5), (9), and (10) respectively, and the Schottky barrier height at the collector is assumed to be 0.9 eV. 

We now discuss the results of model calculations. Fig. 1(a) presents the energy distribution of the hot electrons at zero temperature, and Fig. 1(b) displays the how many hot electrons can overcome the Schottky barrier height at the collector side due to finite temperature effect. At zero temperature, only half of the injected electrons have higher energies than the collector barrier height, and can contribute to the collector current as it is expected. With increasing temperature T the number of electrons having higher energies than the barrier height is increasing, so that the collector current will be also increasing. For instance if there is no event to reduce the current such as angle dependence at the interface \cite{angle}, various scattering in any layer, then the current will be increased roughly 50 $\%$ at room temperature compared with that of zero temperature case.      

Fig. 2 presents the parallel and anti-parallel collector expressed in Eqs. (9) and (10) at finite temperatures. One can clearly see that the parallel and anti-parallel collector current behave differently at finite temperatures. The parallel collector current is increasing up to near the 200 K, and starts to decrease while the anti-parallel current is increasing up to room temperature. One should note the role of two factors such as the hot electron spin polarization and the energy distribution distribution of injected electrons at finite temperatures. As we can see from the Fig. 1, the number of hot electrons are increasing with temperature T we can therefore understand the increasing of both the parallel and anti-parallel collector current. However, roughly speaking, beyond 200 K the parallel and anti-parallel collector current behave differently. We interpret this terms of hot electron spin polarization. Since $1-P_{H_i}(\tilde{\epsilon}(T^\prime)$ and $1+P_{H_i}(\tilde{\epsilon}(T^\prime)$ in Eqs. (9) and (10) contribute to the collector current in the opposite way, for instance the $1-P_{H_i}(\tilde{\epsilon}(T^\prime)$ is increasing with temperature T while the $1+P_{H_i}(\tilde{\epsilon}(T^\prime)$ is decreasing with T, thus they are competing each other and contributing differently to the collector current. As a result, the hot electron spin polarization is contributing to suppress the parallel collector current  and enhance the anti-parallel collector current. Finally the competition between the hot electron spin polarization and the redistribution of hot electron energy due to finite temperature effect controls the collector current at finite temperatures. From the parallel collector current we can note that the hot electron spin polarization has a substantial influence on the collector current above the 200 K since the collector current is expected to be increased roughly 50 $\%$ at room temperature from the Fig. 1 if the hot electron spin polarization has marginal influence on the collector current. Similarly, one can expect the same behavior for the anti-parallel collector current. However, the anti-parallel collector current is increased more than it is expected from the results of the Fig. 1. The overall behaviors of the parallel and anti-parallel collector current show the same trend as in the Fig. 2 of Ref. \cite{Jansen}. If one notes the magnitude of the collector current in Fig. 2, one can realize that the output collector current is roughly $10^{-3}$ times smaller than input current. Since the experimental data shows \cite{Jansen} that the output collector current is approximately reduced by $10^{-6}$ times, then this calculations suggest that the collector current will be reduced by order of 3 in the magnitude. Interestingly, the transfer ratio measurement of non-magnetic transistor \cite{width} shows roughly $10^{-3}$ reduction in the output collector current.       

We now discuss the magnetocurrent. It will be of interest to explore the hot electron spin polarization dependence of magnetocurrent. We therefore display the magnetocurrent with two different cases in Fig. 3. The asterisk represents the magnetocurrent with  $P_{H_i}(\tilde{\epsilon}(T^\prime)= P_{H_i}(\tilde{\epsilon}(0))(1-[T/T_c]^{3/2})$, and the circle is the case for $P_{H_i}(\tilde{\epsilon}(T^\prime)= P_{H_i}(\tilde{\epsilon}(0))(1-[T/T_c]^2)$. In both cases, the magnetocurrent is decreasing with temperature T monotonically, and accords with the experimental data of Ref. \cite{Jansen} in the semi-quantitative manner. One can also note that the magnetocurrent is very sensitive to the form of hot electron spin polarization. 

In conclusion, the hot electron magnetotransport property has been studied theoretically including the influence of the hot electron spin polarization resulting from spin dependent self energy in the ferromagnets and the energy redistribution of hot electrons due to spatial inhomogeneity of Schottky barrier heights at finite temperatures. We obtain that the parallel, antiparallel collector current, and magnetocurrent agree with the experimental data semi-quantitatively. Hence, we suggest that for quantitative understanding the hot electron magnetotransport, it is essential to explore the temperature and energy dependence of the inelastic scattering precesses in the ferromagnetic layer as well as in the normal metal layer, we hope that this work will stimulate to study further related issues mentioned above theoretically and experimentally. 

I would like to thank Dr. P.S. Anil Kumar for useful discussions and comments. 

\newpage
\begin{figure}
\caption{(A) The normalized energy distribution of hot electrons at zero temperature. (B) The number of electrons whicn can contribute to the collector current due to finite temperature effect.}
\end{figure}
\begin{figure}
\caption{The parallel and anti-parallel collector current with the hot electron spin polarization $P_{H_i}(\tilde{\epsilon}(T)= P_{H_i}(\tilde{\epsilon}(0))(1-[T/T_c]^{3/2})$. The critical temperature has been taken as 1200K for Fe and 630 K for Ni.}
\end{figure}
\begin{figure}
\caption{The magnetocurrent at finite temperatures. The asterisk displays the case with $P_{H_i}(\tilde{\epsilon}(T)= P_{H_i}(\tilde{\epsilon}(0))(1-[T/T_c]^{3/2})$, and the circle is for the $P_{H_i}(\tilde{\epsilon}(T)= P_{H_i}(\tilde{\epsilon}(0))(1-[T/T_c]^2)$.}
\end{figure}
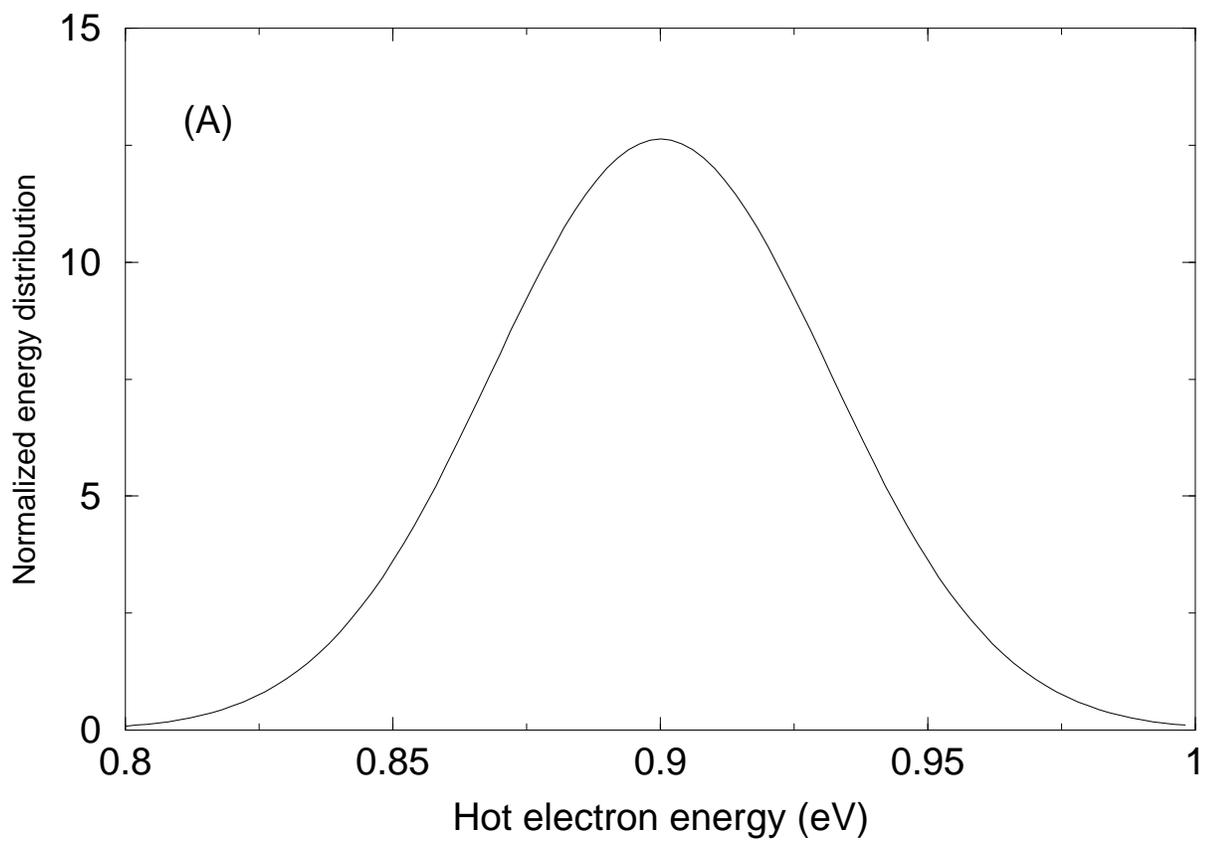
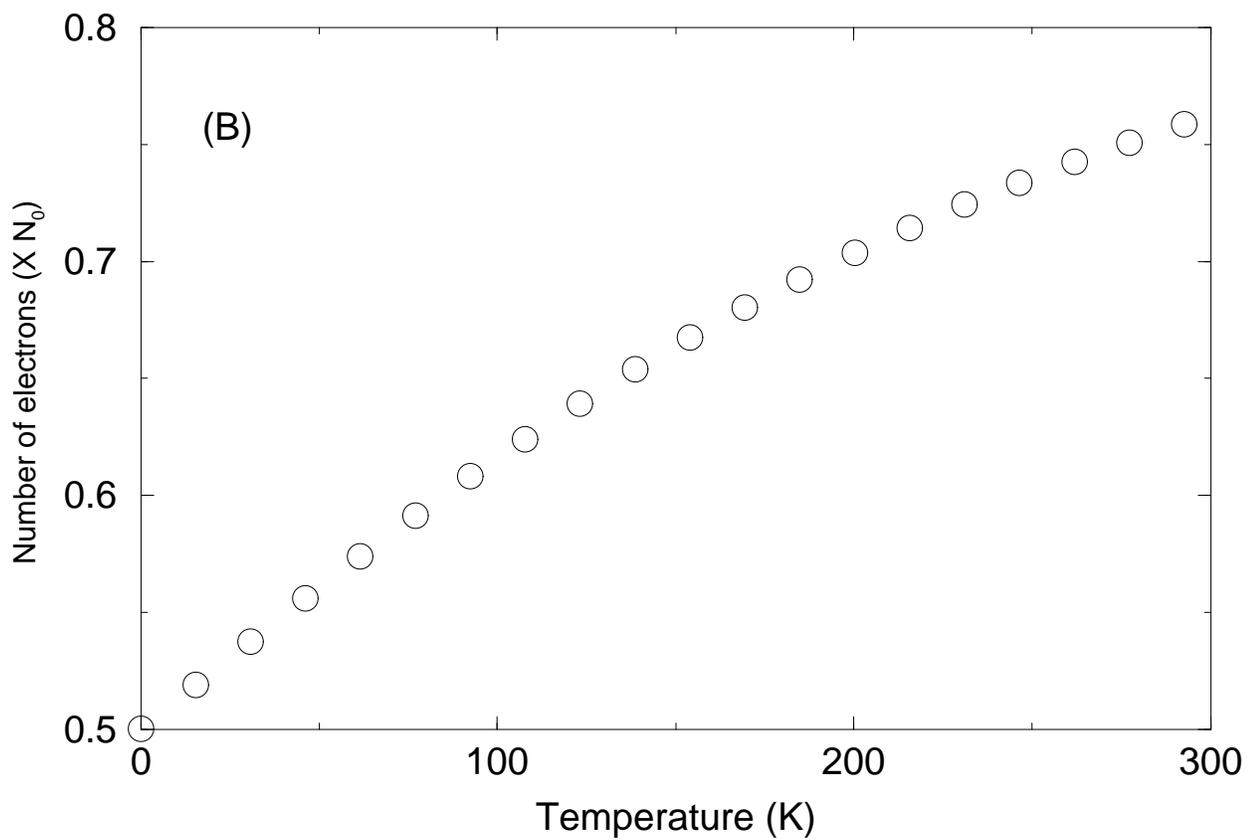
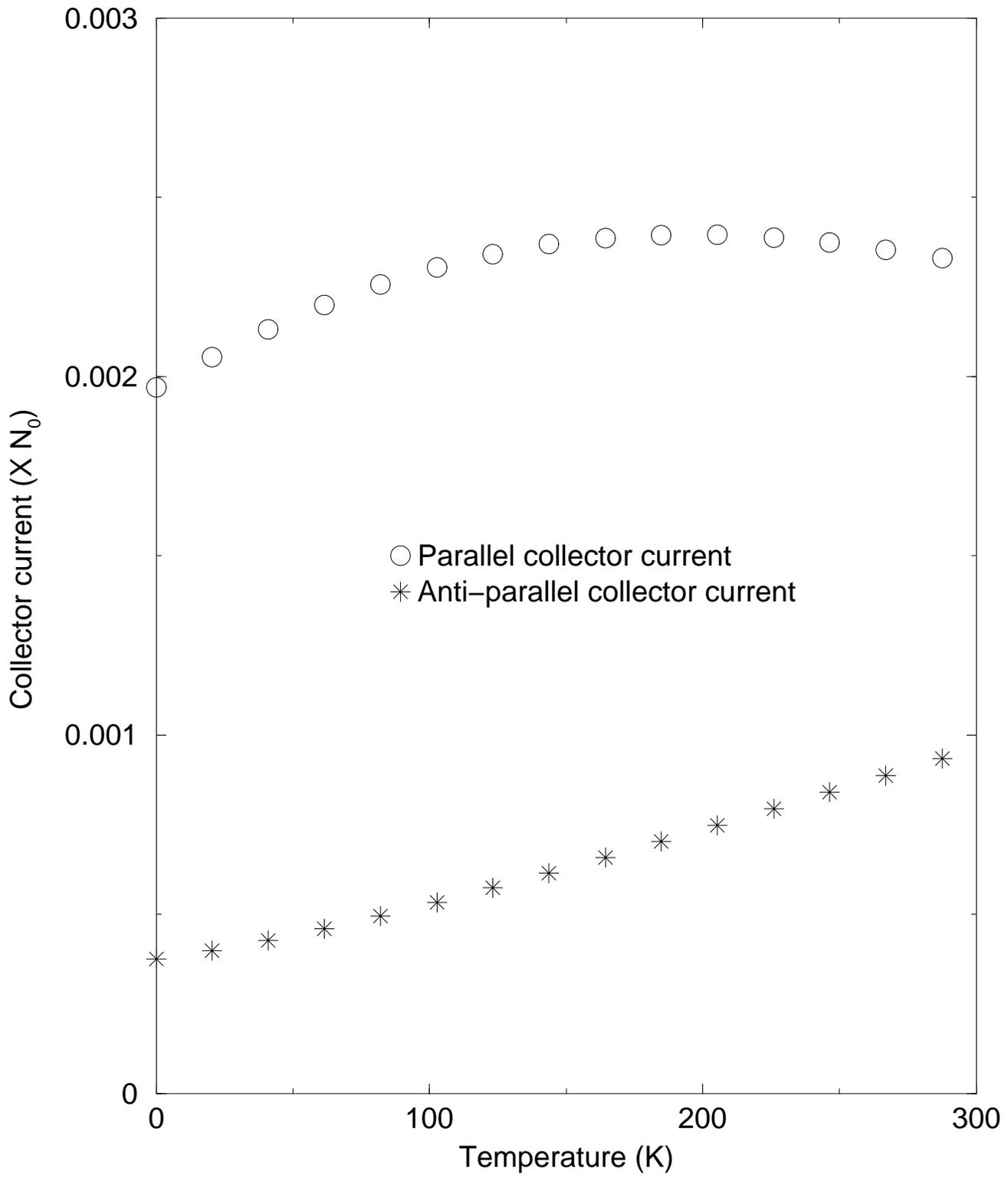
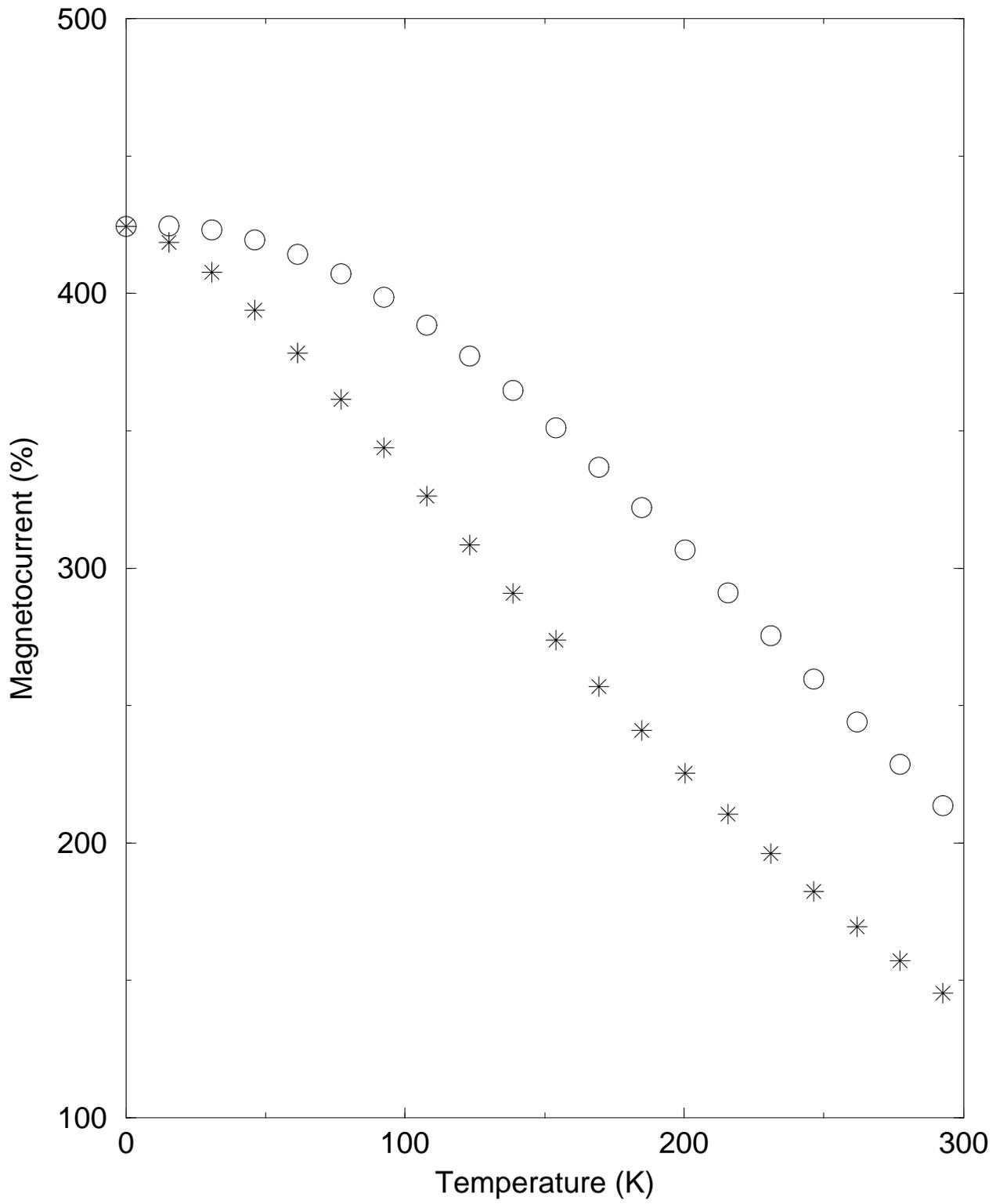
\end{document}